\documentclass{article}

\usepackage{arxiv}

\usepackage[utf8]{inputenc} 
\usepackage[T1]{fontenc}    
\usepackage{hyperref}       
\usepackage{url}            
\usepackage{booktabs}       
\usepackage{amsfonts}       
\usepackage{nicefrac}       
\usepackage{microtype}      
\usepackage{lipsum}
\usepackage{graphicx}
\graphicspath{ {./images/} }

\usepackage{listings}
\title{pixelLOG: Logging of Online Gameplay \\ for Cognitive Research}

\author{
 Zeyu Lu \\
  Department of Computer Science and Engineering\\
  Washington University in St. Louis\\
  1 Brookings Drive\\
  St. Louis, MO 63130\\
  \texttt{mark.lu@wustl.edu}\\
   \And
 Dennis L. Barbour \\
  Department of Biomedical Engineering\\
  Washington University in St. Louis\\
  1 Brookings Drive\\
  St. Louis, MO 63130\\
  \texttt{dbarbour@wustl.edu} \\
}

\begin{document}
\maketitle
\begin{abstract}
Traditional cognitive assessments often rely on isolated, output-focused measurements that may fail to capture the complexity of human cognition in naturalistic settings. We present pixelLOG, a high-performance data collection framework for Spigot-based Minecraft servers designed specifically for process-based cognitive research. Unlike existing frameworks tailored only for artificial intelligence agents, pixelLOG also enables human behavioral tracking in multi-player/multi-agent environments. Operating at configurable frequencies up to and exceeding 20 updates per second, the system captures comprehensive behavioral data through a hybrid approach of active state polling and passive event monitoring. By leveraging Spigot’s extensible API, pixelLOG facilitates robust session isolation and produces structured JSON outputs integrable with standard analytical pipelines. This framework bridges the gap between decontextualized laboratory assessments and richer, more ecologically valid tasks, enabling high-resolution analysis of cognitive processes as they unfold in complex, virtual environments.
\end{abstract}

\keywords{Cognitive assessment \and behavioral data collection \and process-based assessment \and Minecraft \and Spigot plugin \and agentic AI}

\section{Introduction}
Cognitive assessment methodologies have historically been constrained by laboratory-like conditions and narrowly defined tasks, often emphasizing the measurement of singular executive functions—such as working memory, inhibitory control, or attention—through static, outcome-based metrics \cite{miyake_unity_2000}. While these conventional approaches have contributed to our foundational understanding of cognition, they frequently lack ecological validity \cite{hedge_reliability_2018}, failing to capture the complex, context-dependent, and adaptive nature of human cognitive processes as they unfold in multifaceted, real-world environments. Moreover, recent evidence suggests that many well-established executive function tasks may measure little more than individual differences in the speed of information uptake rather than tapping into distinct cognitive constructs \cite{loffler_common_2024}. Such restrictions limit our ability to observe how individuals integrate various cognitive strategies, respond dynamically to changing scenarios, or leverage available resources to achieve their goals.

Increasingly, scholars and practitioners have sought assessment frameworks that provide process-oriented, fine-grained behavioral data, thus transcending the conventional confines of controlled laboratory tasks. In this evolving landscape, virtual platforms and game-based assessments have garnered attention for their capacity to offer rich, interactive contexts that more closely approximate everyday cognitive challenges. Among such platforms, Minecraft—a widely adopted, open-ended sandbox environment—distinguishes itself by enabling the simulation of complex tasks requiring navigation, resource management, problem-solving, and creativity. Its research potential is further augmented by Spigot, an open-source server modification platform that empowers researchers to create customized plugins, alter game mechanics, and collect detailed player data without compromising performance.

We introduce \emph{pixelLOG (Logging of Online Gameplay)}\footnote{Source code is available at \url{https://github.com/NMIL230/nmil-p-mc-pixelLOG}}, a high-performance, plugin-based logging framework that integrates seamlessly with the Spigot modification layer for Minecraft servers. In contrast to many existing cognitive research tools that rely on predetermined scenarios and coarse data sampling, pixelLOG enables fine-grained, player-level data acquisition at flexible frequencies exceeding 20 updates per second. By capturing a broad range of behavioral and environmental variables—from player health and location to event-based triggers such as block placements, item usage, or entity interactions—pixelLOG transcends the constraints of purely output-focused methods. It thus supports a process-based, granular examination of how individuals engage with, adapt to, and evolve their strategies within complex, multi-dimensional tasks.

The pixelLOG architecture not only facilitates the precise mapping of individual cognitive trajectories onto rich environmental cues but also enables concurrent multi-player/multi-agent data collection, supporting large-scale comparative studies and interactive/collective research with robust session-based analytics. The structured JSON outputs that pixelLOG generates integrate seamlessly into standard computational pipelines, thereby allowing advanced statistical, machine learning, and visualization techniques to uncover latent patterns and cognitive strategies over time.
In sum, pixelLOG addresses the critical need for richer, more ecologically valid assessment methodologies by offering:

\begin{itemize}
    \item \textbf{Fine-Grained Data Acquisition:} A highly customizable logging system capable of capturing a wide range of data points—environmental, behavioral, and cognitive—at user-defined frequencies.

    \item \textbf{Scalability and Adaptability:} A modular, scalable architecture supporting multi-player contexts, session management, and selective data acquisition suited for diverse experimental designs.

    \item \textbf{Structured and Extensible Outputs:} Organized, JSON-formatted data readily compatible with advanced analytical tools, thereby streamlining subsequent inference, modeling, and visualization efforts.
\end{itemize}

Through these contributions, pixelLOG advances the field of cognitive assessment into more naturalistic, dynamic domains, expanding the boundaries of how we measure, understand, and ultimately enhance human cognition in real-world or near-real-world contexts.

\section{Related work}
\label{sec:headings}
A variety of platforms and research tools, such as Microsoft's Project Malmo \cite{perez-liebana_multi-agent_2025} and the MineDojo framework \cite{fan_minedojo_2022}, have emerged to facilitate experimentation, data collection, and reinforcement learning (RL) research within Minecraft \cite{wang_voyager_2023, qin_mp5_2024, wang_jarvis-1_2024, hafner_mastering_2025}. These frameworks provide standardized interfaces for agent interaction and observation in controlled environments. While these agent-centric RL studies collect in-game data as observation space, they typically lack the high-precision, high-frequency, and configurable data collection pipelines necessary for human cognitive research. Moreover, these experimental platforms often operate in isolation, with limited extensibility and customization capabilities.

In contrast, pixelLOG is specifically designed for researchers investigating human cognitive processes in dynamic virtual environments. By leveraging Spigot's event-driven architecture and implementing a custom plugin-based solution, the system delivers high-frequency polling, granular event capturing, and robust per-player data isolation. Unlike existing solutions that may impose constraints on data granularity or system extensibility, pixelLOG's modular architecture provides the flexibility and performance required for comprehensive cognitive research. This utility has been demonstrated in recent applications: the framework served as the data collection backbone for \emph{pixelDOPA (Digital Online Psychometric Assessment)}, enabling the validation of immersive cognitive minigames against the NIH Toolbox \cite{marticorena_immersive_2025}, and supported real-time data integration for \emph{AMLEC}, a multidimensional Bayesian active machine learning study of working memory \cite{marticorena_multidimensional_2025}.

\section{Methods}

\begin{figure}
  \centering
  \includegraphics[width=\textwidth]{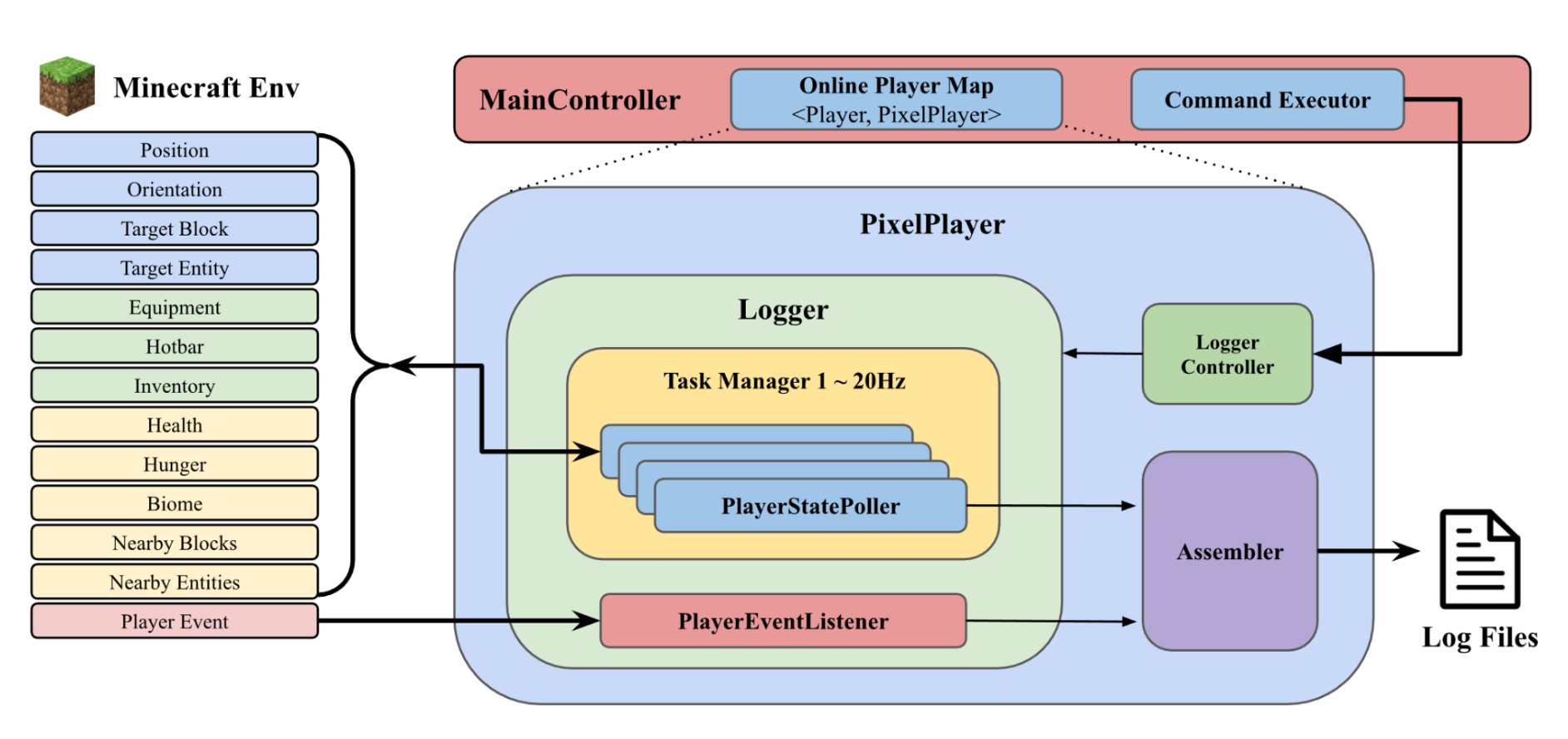}
  \caption{pixelLOG consists of several key components, each designed with specific responsibilities to ensure efficient data collection and processing.}
  \label{fig:diagram}
\end{figure}

The pixelLOG system is conceived as a modular and extensible data collection framework designed to operate within a multiplayer Minecraft environment. As shown in Figure \ref{fig:diagram}, the architecture consists of distinct layers responsible for player management, data acquisition, event interception, and structured output generation. By dividing responsibilities across modular components, the system facilitates scalability, maintainability, and precise adaptation to varying research requirements. This layered arrangement ensures that the fundamental data-handling logic remains loosely coupled, thus enabling incremental enhancements or the integration of advanced analytical modules with minimal disruption.

\subsection{Scalable multi-player data management}

In multiplayer environments, robust isolation of behavioral data collection across participants is essential. pixelLOG implements this solution through a hierarchical architecture centered on the MainController, which interfaces with Spigot's core systems. Upon player connection, the MainController instantiates a dedicated PixelPlayer object that encapsulates all participant-specific data collection processes. This architecture ensures linear scalability with increasing concurrent player counts while maintaining system stability.

Each PixelPlayer instance orchestrates an isolated suite of data collection modules and logging components. This modular design prevents data cross-contamination, ensuring that behavioral patterns and environmental observations remain discretely partitioned between participants. While initial implementations utilized centralized logging mechanisms, empirical testing revealed susceptibility to bottlenecks and intermittent data loss under high load conditions. The transition to distributed, thread-safe queues for individual players has demonstrably enhanced both collection reliability and system performance under peak utilization scenarios, while maintaining strict data isolation guarantees.

Additionally, as the primary plugin class, the MainController manages essential Spigot plugin lifecycle operations, including plugin initialization and finalization, event listener registration, command registration, and configuration management. This centralized control ensures proper resource allocation and cleanup throughout the plugin's operational lifecycle.

\subsection{Adaptive temporal resolution in data collection}

Comprehensive analysis often requires data captured at differing temporal granularities. pixelLOG employs a dedicated Logger component that coordinates multiple PlayerStatePollers, each operating at configurable frequencies. During initialization, the Logger establishes distinct scheduled tasks, each executing at user-defined intervals through the Spigot scheduler system. These tasks invoke corresponding PlayerStatePollers and register with the MainController for coordinated operation within the plugin ecosystem.

The system implements a stratified polling strategy: higher-frequency pollers (e.g., 10–20 updates per second) capture rapidly evolving attributes such as player avatar position, orientation, and velocity, while lower-frequency pollers survey relatively static environmental parameters, such as biome types or stationary entities, at intervals of once per second or less. This adaptive temporal resolution enables researchers to obtain high-fidelity data for dynamic phenomena while optimizing computational overhead through reduced sampling rates for stable environmental attributes.

\subsection{Integrating active polling with event-driven data capture}

While scheduled polling tasks ensure a steady stream of state-based data, certain behavioral indicators emerge from discrete events—such as block placements, combat engagements, or item usage. To incorporate these episodic markers, pixelLOG implements a hybrid data collection strategy. During initialization, the Logger instantiates a configurable PlayerEventListener for the player and registers it with the MainController, enabling targeted interception of player-specific game events through the Spigot event system.

By fusing asynchronous event-driven data with continuous polling data, researchers can anchor moment-to-moment behavioral patterns within the broader context of meaningful in-game actions. This dual-method data capture strategy enhances analytical fidelity, enabling comprehensive behavioral models that integrate both continuous state trajectories and pivotal interaction events.

\subsection{Data integration and structured output}

To unify the heterogeneous data streams generated by pollers and event listeners, pixelLOG employs an Assembler component. The Assembler harmonizes all incoming data into chronologically ordered, structured JSON files, ensuring consistency and simplifying subsequent analysis. These formatted outputs can be easily integrated into common data science toolchains, thus minimizing overhead during post-processing and visualization phases. Although current outputs are stored locally, an updated version of pixelLOG includes a configurable Transmission Control Protocol (TCP) module, allowing JSON-based logs to be forwarded directly to external servers or databases for long-term archival and automated analytical workflows.

\section{Conclusion}

pixelLOG presents a novel approach to cognitive research through high-performance behavioral data collection in Minecraft environments. Our framework enables comprehensive capture of both player states and events at configurable frequencies while maintaining system stability. The structured JSON output format facilitates seamless integration with standard analytical pipelines, including statistical packages, machine learning frameworks, and visualization tools, enabling researchers to conduct detailed temporal analyses and identify patterns linking environmental factors to cognitive strategies.

By providing unprecedented access to detailed behavioral and contextual data, pixelLOG bridges the gap between traditional outcome-focused cognitive assessments and richer, process-based evaluations. This work advances cognitive research methodology by enabling the study of cognitive processes in naturalistic, interactive virtual environments, opening new avenues for empirical research and theoretical development in cognitive science.

\section{Acknowledgments}

This study was supported in part by the Washington University OVCR Seed Grant and Here and Next Programs, as well as NSF NAIRR240235.

\bibliographystyle{unsrt}  
\bibliography{zLibraryNMIL}  

@article{hafner_mastering_2025,
	title = {Mastering diverse control tasks through world models},
	volume = {640},
	issn = {1476-4687},
	doi = {10.1038/s41586-025-08744-2},
	language = {eng},
	number = {8059},
	journal = {Nature},
	author = {Hafner, Danijar and Pasukonis, Jurgis and Ba, Jimmy and Lillicrap, Timothy},
	month = apr,
	year = {2025},
	pages = {647--653},
}

@misc{marticorena_immersive_2025,
	title = {Immersive virtual games: {Winners} for deep cognitive assessment},
	shorttitle = {Immersive virtual games},
	url = {http://arxiv.org/abs/2502.10290},
	doi = {10.48550/arXiv.2502.10290},
	urldate = {2025-02-17},
	publisher = {arXiv},
	author = {Marticorena, Dom C. P. and Lu, Zeyu and Wissmann, Chris and Agarwal, Yash and Garrison, David and Zempel, John M. and Barbour, Dennis L.},
	month = feb,
	year = {2025},
	note = {arXiv:2502.10290 [cs]},
}

@article{wang_jarvis-1_2024,
	title = {Jarvis-1: {Open}-world multi-task agents with memory-augmented multimodal language models},
	shorttitle = {Jarvis-1},
	url = {https://ieeexplore.ieee.org/abstract/document/10778628/},
	urldate = {2026-02-06},
	journal = {IEEE Transactions on Pattern Analysis and Machine Intelligence},
	publisher = {IEEE},
	author = {Wang, Zihao and Cai, Shaofei and Liu, Anji and Jin, Yonggang and Hou, Jinbing and Zhang, Bowei and Lin, Haowei and He, Zhaofeng and Zheng, Zilong and Yang, Yaodong},
	year = {2024},
}

@inproceedings{qin_mp5_2024,
	title = {Mp5: {A} multi-modal open-ended embodied system in minecraft via active perception},
	shorttitle = {Mp5},
	url = {https://ieeexplore.ieee.org/abstract/document/10657187/},
	urldate = {2026-02-06},
	booktitle = {2024 {IEEE}/{CVF} {Conference} on {Computer} {Vision} and {Pattern} {Recognition} ({CVPR})},
	publisher = {IEEE},
	author = {Qin, Yiran and Zhou, Enshen and Liu, Qichang and Yin, Zhenfei and Sheng, Lu and Zhang, Ruimao and Qiao, Yu and Shao, Jing},
	year = {2024},
	pages = {16307--16316},
}

@misc{perez-liebana_multi-agent_2025,
	title = {The {Multi}-{Agent} {Reinforcement} {Learning} in {MalmÖ} ({MARLÖ}) {Competition}},
	url = {http://arxiv.org/abs/1901.08129},
	doi = {10.48550/arXiv.1901.08129},
	urldate = {2026-02-06},
	publisher = {arXiv},
	author = {Perez-Liebana, Diego and Hofmann, Katja and Mohanty, Sharada Prasanna and Kuno, Noboru and Kramer, Andre and Devlin, Sam and Gaina, Raluca D. and Ionita, Daniel},
	month = apr,
	year = {2025},
	note = {arXiv:1901.08129 [cs]},
}

@article{hedge_reliability_2018,
	title = {The reliability paradox: {Why} robust cognitive tasks do not produce reliable individual differences},
	volume = {50},
	issn = {1554-3528},
	shorttitle = {The reliability paradox},
	url = {https://doi.org/10.3758/s13428-017-0935-1},
	doi = {10.3758/s13428-017-0935-1},
	language = {en},
	number = {3},
	urldate = {2024-10-14},
	journal = {Behavior Research Methods},
	author = {Hedge, Craig and Powell, Georgina and Sumner, Petroc},
	month = jun,
	year = {2018},
	pages = {1166--1186},
}

@misc{marticorena_multidimensional_2025,
	title = {Multidimensional {Bayesian} active machine learning of working memory task performance},
	url = {http://arxiv.org/abs/2510.00375},
	doi = {10.48550/arXiv.2510.00375},
	urldate = {2025-10-02},
	publisher = {arXiv},
	author = {Marticorena, Dom C. P. and Wissmann, Chris and Lu, Zeyu and Barbour, Dennis L.},
	month = oct,
	year = {2025},
	note = {arXiv:2510.00375 [cs]},
}

@misc{fan_minedojo_2022,
	title = {{MineDojo}: {Building} {Open}-{Ended} {Embodied} {Agents} with {Internet}-{Scale} {Knowledge}},
	shorttitle = {{MineDojo}},
	url = {http://arxiv.org/abs/2206.08853},
	doi = {10.48550/arXiv.2206.08853},
	urldate = {2025-03-18},
	publisher = {arXiv},
	author = {Fan, Linxi and Wang, Guanzhi and Jiang, Yunfan and Mandlekar, Ajay and Yang, Yuncong and Zhu, Haoyi and Tang, Andrew and Huang, De-An and Zhu, Yuke and Anandkumar, Anima},
	month = nov,
	year = {2022},
	note = {arXiv:2206.08853 [cs]},
}

@article{loffler_common_2024,
	title = {The common factor of executive functions measures nothing but speed of information uptake},
	volume = {88},
	issn = {1430-2772},
	url = {https://doi.org/10.1007/s00426-023-01924-7},
	doi = {10.1007/s00426-023-01924-7},
	language = {en},
	number = {4},
	urldate = {2024-10-14},
	journal = {Psychological Research},
	author = {Löffler, Christoph and Frischkorn, Gidon T. and Hagemann, Dirk and Sadus, Kathrin and Schubert, Anna-Lena},
	month = jun,
	year = {2024},
	pages = {1092--1114},
}

@misc{wang_voyager_2023,
	title = {Voyager: {An} {Open}-{Ended} {Embodied} {Agent} with {Large} {Language} {Models}},
	shorttitle = {Voyager},
	url = {http://arxiv.org/abs/2305.16291},
	urldate = {2024-01-22},
	publisher = {arXiv},
	author = {Wang, Guanzhi and Xie, Yuqi and Jiang, Yunfan and Mandlekar, Ajay and Xiao, Chaowei and Zhu, Yuke and Fan, Linxi and Anandkumar, Anima},
	month = oct,
	year = {2023},
	note = {arXiv:2305.16291 [cs]},
}

@article{miyake_unity_2000,
	title = {The unity and diversity of executive functions and their contributions to complex "{Frontal} {Lobe}" tasks: a latent variable analysis},
	volume = {41},
	issn = {0010-0285},
	shorttitle = {The unity and diversity of executive functions and their contributions to complex "{Frontal} {Lobe}" tasks},
	doi = {10.1006/cogp.1999.0734},
	language = {eng},
	number = {1},
	journal = {Cognitive Psychology},
	author = {Miyake, A. and Friedman, N. P. and Emerson, M. J. and Witzki, A. H. and Howerter, A. and Wager, T. D.},
	month = aug,
	year = {2000},
	pages = {49--100},
}






\section{Appendix}

\subsection{Installation}

\begin{itemize}
  \item \textbf{Prerequisites}
  \begin{itemize}
    \item A Spigot Server running version 1.20.4.
    \item Java Development Kit (JDK) 21 (if building from source).
    \end{itemize}
  \item \textbf{Option A: Using Pre-built Release}
  \begin{itemize}
      \item Download the latest release of pixelLOG from the Releases page.
      \item Place the .jar file into the plugins directory of your Spigot server.
    \end{itemize}
  \item \textbf{Option B: Building from Source (Maven)}
    \begin{itemize}
      \item Clone the repository from GitHub: \url{https://github.com/NMIL230/nmil-p-mc-pixelLOG}
      \item Run the Maven \texttt{package} command in the project root directory
      \item Locate the compiled jar file at \texttt{target/pixelLOG.jar}.
      \item Move this file to your server's plugins directory.
  \end{itemize}
\end{itemize}

\subsection{Initialization}

\begin{itemize}
  \item Start the Spigot server by executing the following command: \texttt{./start.sh}
  \item After initialization, inspect the server logs to verify successful plugin activation. Upon proper loading, the server logs will display: pixelLOG Enabled
  \item If this confirmation message is present in your server logs, pixelLOG is now active and ready to collect player data.
\end{itemize}

\subsection{Usage}

pixelLOG provides a set of administrative commands enabling authorized users to initiate and terminate logging sessions, as well as to query the current plugin version. 

\subsubsection{Available commands}

  \begin{tabular}{lp{2.5in}l}
    \hline
    Command      & Description                                                        & Usage                     \\
    \hline
    \texttt{pl-start}    & Initiates data logging for the player issuing the command. & \texttt{/pl-start}        \\
    \texttt{pl-stop}     & Terminates data logging for the player issuing the command.& \texttt{/pl-stop}         \\
    \texttt{pl-start-op} & Initiates data logging for a specified online player, requires operator (OP) privileges. & \texttt{/pl-start-op [username]} \\
    \texttt{pl-stop-op}  & Terminates data logging for a specified online player, requires operator (OP) privileges.& \texttt{/pl-stop-op [username]}  \\
    \texttt{pl-version}  & Displays the current pixelLOG plugin version (primarily for diagnostic purposes). & \texttt{/pl-version}      \\
    \hline
  \end{tabular}

\subsubsection{Log file generation}

When data collection ceases---either through a stop command or when the player exits the game---pixelLOG produces a structured JSON logfile. These files are stored in the \texttt{/PixelLogs} directory located within your primary Spigot server folder.

\subsection{Data collection reference}

pixelLOG generates structured JSON datasets that integrate continuous state measurements with discrete event occurrences, facilitating temporal analysis of player behaviors. While implementation details are covered in Appendix A, this section defines the data organization, schema structures, and specific collection parameter

\subsubsection{Log structure and data organization}

pixelLOG organizes collected data into structured JSON files, with each file representing a complete player session. The log structure follows a hierarchical format designed for efficient data retrieval and analysis. A representative configuration of the log structure is illustrated below:

The root level contains session metadata, including unique identifiers, player information, temporal boundaries, and operational parameters:

\begin{lstlisting}
{
    "logfile_id": "Unique_Session_Identifier",
    "filename": "Player_UTCTimeString",
    "username": "Player_ID",
    "game_start_time": "Session_Start_UTCTimeString",
    "game_end_time": "Session_End_UTCTimeString",
    "plugin_version": "Version_String",
}

\end{lstlisting}

The core behavioral data is stored in a chronologically ordered array of log entries, each categorized by type and containing detailed contextual information. Player state data can be collected by PlayerStatePoller at various frequencies, with the following example representing data captured at a rate of 20 updates per second:

\begin{lstlisting}
{
    "type": "HIGH_FREQUENCY_LOG_20Hz",
    "time": "UTCTimeString",
    "game_tick": "Server_Tick_Count",
    "location": {"x": 0.0, "y": 0.0, "z": 0.0},
    "view": {"pitch": 0.0, "yaw": 0.0},
    "ray_tracing_block": {
        "hit_location": "coordinates",
        "block_type": "material"
    }
}
\end{lstlisting}

Event logs capture discrete player interactions through the PlayerEventListener, with each log entry containing event-specific data and contextual information. A basic example of an event log structure is as follows:

\begin{lstlisting}
{
    "type": "EVENT_LOG",
    "time": "UTCTimeString",
    "game_tick": "Server_Tick_Count",
    "event": "Event_Name"
    "event_info": {
        // This section contains event-specific information
    }
}
\end{lstlisting}

This structured format represents one possible configuration of pixelLOG's data collection capabilities. The complete range of available data types and collection parameters is detailed in the subsequent section on data collection specifications.

\subsubsection{Data collection specifications}

pixelLOG implements comprehensive data collection across two primary categories, player states and events. We support the following data types:

\begin{itemize}
  \item \textbf{Player state variables}
  \begin{itemize}
    \item Location \& View: Precise spatial data (x, y, z), combined with orientation metrics (pitch, yaw), allow for reconstructing navigation paths and identifying exploration patterns, spatial memory use, and attentional focus.
    \item Health \& Hunger: Physiological metrics that can correlate with task difficulty, resource availability in player performance.
    \item Ray-Traced Target (Blocks \& Entities): Real-time tracking of player's line-of-sight interactions, providing precise hit locations and types for both blocks and entities the player is targeting.
    \item Nearby Blocks, Entities and Biome: Continuous assessment of surrounding blocks, entities within a configurable radius, and biome classification, establishing the immediate context of player actions.
  \end{itemize}
  \item{Player events}
  \begin{itemize}
    \item Block Events: Records player interactions with the environment, including block breaking, placement, and damage initiation. Each event captures the block type, location coordinates, and relevant tool information.
    \item Interaction Events: Documents all entity-related interactions, including direct engagement (e.g., shearing, feeding), combat scenarios, and interactions with functional blocks (e.g., crafting tables, furnaces, chests). Events store entity type, interaction type, block type when applicable, and additional context-specific data.
    \item Combat Events: Tracks combat-related activities, recording both offensive actions (player attacking entities) and defensive scenarios (player receiving damage). Each event includes damage values, damage cause, and involved entity information.
    \item Item Events: Monitors item-related activities including consumption, equipment changes, inventory manipulation, and crafting operations. Events store detailed information about items involved, quantities, and specific inventory slots when applicable.
  \end{itemize}
\end{itemize}

\subsubsection{Player state data collection}

pixelLOG supports comprehensive player state data collection at configurable frequencies. The following data points can be collected:

\begin{itemize}
  \item{Basic Player States}
  \begin{itemize}
    \item Health: Current player health level
    \item Hunger: Current hunger level
    \item Location: World coordinates (x, y, z)
    \item View: Player orientation (yaw and pitch)
    \item Environmental Interaction
    \item Target Block: Block currently targeted by player's cursor
    \item Nearby Entities: All entities within a defined radius
    \item Nearby Blocks: Blocks within specified observation radius
    \item Biome: Current location's biome type
  \end{itemize}
  \item Inventory and Equipment
    \begin{itemize}
    \item Hot Bar: Items in quick access toolbar
    \item Inventory: Complete inventory contents
    \item Equipment: Worn armor and held items (main/off-hand)
    \item Advanced Tracking
    \item Ray Trace Block: Block targeted by player's line of sight
    \item Ray Trace Entities: Entities in player's line of sight
  \end{itemize}
\end{itemize}

\subsubsection{Event logging system}

The system captures the following player-triggered events:

\paragraph{Block interaction events}

\begin{itemize}
  \item Block Break Event
  \begin{itemize}
    \item Trigger: Player breaks a block
    \item Captured Data: player, event, block\_type, block\_location
  \end{itemize}  
  \item Block Damage Event
  \begin{itemize}
    \item Trigger: Player begins mining a block
    \item Captured Data: player, event, block\_type, block\_location, item\_in\_hand
  \end{itemize}  
  \item Block Place Event
  \begin{itemize} 
    \item Trigger: Player places a block
    \item Captured Data: player, event, block\_type, block\_location
  \end{itemize}
\end{itemize}

\paragraph{Entity interaction events}

\begin{itemize}
  \item Player Interact Event
  \begin{itemize}
    \item Trigger: Player interacts with game world
    \item Captured Data: player, event, action, item\_in\_hand, block\_type, block\_location
  \end{itemize}
  \item Entity Interaction Events
  \begin{itemize}
    \item Player-Entity Interaction
    \begin{itemize}
      \item Trigger: Direct entity interaction
      \item Captured Data: player, event, entity\_type
    \end{itemize}
    \item Entity Damage Events
    \begin{itemize}
      \item Trigger: Combat or damage interactions
      \item Captured Data: player, event, entity\_type, damage
    \end{itemize}
  \end{itemize}
\end{itemize}

\paragraph{Item management events}

\begin{itemize} 
  \item Item Usage Events
  \begin{itemize} 
    \item Consumption Event
    \begin{itemize} 
      \item Trigger: Item consumption
      \item Captured Data: player, event, item
    \end{itemize} 
  \item Item Holding Event
    \begin{itemize} 
      \item Trigger: Hotbar item change
      \item Captured Data: player, event, new\_item
    \end{itemize} 
  \end{itemize} 
  \item Item Transfer Events
  \begin{itemize} 
    \item Drop Item Event
    \begin{itemize} 
      \item Trigger: Item dropping
      \item Captured Data: player, event, item\_type, amount
    \end{itemize} 
    \item Pickup Item Event
    \begin{itemize} 
      \item Trigger: Item collection
      \item Captured Data: player, event, item\_type, amount
    \end{itemize} 
  \end{itemize} 
  \item Inventory Management
  \begin{itemize} 
    \item Inventory Click Event
    \begin{itemize} 
      \item Trigger: Inventory interaction
      \item Captured Data: player, event, slot, clicked\_item
    \end{itemize} 
    \item Crafting Event
    \begin{itemize} 
      \item Trigger: Item crafting
      \item Captured Data: player, event, crafted\_item, amount
    \end{itemize} 
  \end{itemize} 
\end{itemize} 

\end{document}